\documentclass{mn2e}
\usepackage[dvips]{graphicx}

\begin{document}

   \title[Bulge-disc connection]{The Effects of Mergers on the Formation 
of Disc-Bulge Systems in Hierarchical Clustering Scenarios}

   \author[C. Scannapieco and P.B. Tissera]
{C. Scannapieco$^1$\thanks{E-mail: cecilia@iafe.uba.ar}
and P.B. Tissera$^{1,2}$\\
  $^1$Instituto de Astronom\'{\i}a y F\'{\i}sica del Espacio, I.A.F.E.,
 Casilla de Correos  67, Suc.\ 28, Buenos
  Aires, 1428, Argentina\\
  $^2$Consejo Nacional de Investigaciones Cient\'\i ficas y T\'ecnicas,
CONICET   }

   \maketitle

   \begin{abstract}

We study the effects of mergers on the structural properties of
disc-like systems by using Smooth Particle Hydrodynamical (SPH) 
numerical simulations in hierarchical
clustering scenarios. In order to assess the effects of mergers
on the mass distributions
we performed a bulge-disc decomposition of the projected surface density
of the systems at different stages of the merger process.
We  assumed an exponential law for the disc component
and the S\'ersic law for the bulges. We found that simulated objects at
$z=0$ have bulge profiles with shape parameters $n\approx 1$,
 consistent with observational results of spiral galaxies.
The complete sample of simulated objects at $z=0$ and $z>0$ shows that $n$
takes values in the range $n\approx 0.4 - 4$.
We found that secular evolution tends to produce exponential bulge
profiles, while the fusion of baryonic cores tends to increase the
$n$ value and helps to generate the
correlation between $B/D$ and $n$. We found no dependence on the relative
mass of the colliding objects. Our results suggest that  mergers,
 through secular evolution and fusions, could produce the transformation
of galactic objects along the Hubble sequence by driving a morphological
loop that might also depend  on the properties of the 
central galactic potential wells, which are also affected by mergers.

   \end{abstract}


\section{Introduction}
 
The origin of the Hubble sequence is still a controversial issue. 
In particular, spiral galaxies seem to evolve along this
phenomenological classification although the  physical mechanisms
behind these morphological changes are not fully understood.
The structural parameters, the gas abundances and the star formation
activity vary along galaxies in the Hubble sequence in the sense
that disc-dominated systems are also the more gaseous ones,
 experiencing on-going star formation. The large database of
galaxies  gathered in the last years have allowed to get more detailed
information on the properties of different morphological type
galaxies in the local Universe.

Observational results show that when a double
exponential  decomposition
is applied to the sufarce luminosity density of late-type spirals,
 the scalelengths of the bulge and disc components
seem to be restricted to a certain value $<r_{\rm b}/r_{\rm d}> \approx 0.10$
(Courteau, de Jong \& Broeils 1996, hereafter CdJB96). 
This restriction in the range of possible
scalelengths for spirals has been interpreted as 
a proof for secular evolution to be the responsible mechanism
for bulge formation from an already in place disc structure.
Recently, MacArthur, Courteau \& Holtzman (2002, hereafter
MCH02) studied 
a larger sample of late-type spirals and carried out a bulge-disc decomposition
assuming a S\'ersic law for the bulge surface luminosity density.
These authors found a continuous distribution
of the shape parameter of the bulge ($n\approx 0.2-2$)
 with a  maximum at $n \approx 0.90$.
They also found a restricted
range of values for the scalelengths of the bulge and disc when
the shape parameter is $n \approx 1$,
$<r_{\rm b}/r_{\rm d}>=0.13 \pm 0.06$ 
in the $R$-band. 
Previous works have also found that bulges of spirals
can be fitted by a S\'ersic law with different shape parameters
(e.g., Andredakis, Peletier \& Ballcels 1995, hereafter APB95; 
Khosroshahi, Wadadekar \& Kembhavi 2000,  hereafter KWK00;
Graham \& de Block 2001,  hereafter GdB01). APB95
and GdB01 also found a correlation between the luminosity
bulge-to-disc ratio $B/D$ and the shape parameter of the bulge, 
which also correlates with morphological type.
These observational results suggest a possible connection between 
the formation of the bulge and disc components. 
They also support the idea  that $n$ could 
be used as a good indicator of position in the Hubble sequence.

Several theories have been developed to explain the formation
of the disc and bulge components. 
In the case of the disc systems, the
 standard model is based on three hypothesis: 
the angular momentum is acquired through cosmological
torques (Peebles 1969), baryons and dark matter have the same
specific angular momentum content ($J$), and this specific angular momentum 
is conserved during the collapse and cooling of baryons (Fall \& 
Efstathiou 1980,  hereafter FE80).
This model has been successful in reproducing several
observational results in analytical and semi-analytical models 
(e.g., Dalcanton, Spergel \& Summers 1997;  Mo, Mao \& White 1998).
However, serious problems arose in numerical simulations 
where an important angular momentum
transfer from baryons to the dark matter haloes during mergers was detected,
breaking the condition of $J$ conservation.
Dom\'{\i}nguez-Tenreiro, Tissera \& S\'aiz (1998,  hereafter DT98)
showed that  a disc-like structure with observational counterpart
 (S\'aiz et al. 2001,  hereafter S01) can be built up 
 if a compact stellar bulge is allowed to 
form without depleting the gas reservoir of the system.
 These stellar bulges provide stability to the gaseous disc systems
which are capable of conserving a non-negligible fraction of its angular
momentum during violent events.
Supernova energy feedback could also contribute to the formation of the
disc component, regulating the star formation rate and preventing early
catastrophic depletion of the gas into stars.
 Probably both mechanisms, the formation of a compact stellar
bulge which assures the axisymmetrical character of the potential well
and energy feedback, work together in nature to allow the formation
of  spiral galaxies (e.g. Weil, Eke \& Efstathiou 1998). 

The formation of bulges is a more complex task  since
several mechanisms could be acting together
 such as monolithic collapse
(Gilmore \& Wyse 1998), mergers (Kauffman, Guiderdoni \& White
1994) and
secular evolution (Pfenniger \& Norman 1990).
In general, analytical models and pre-prepared
simulations have focused in one or two of them at the time 
(e.g., van den Bosch 1999; Aguerri, Balcells \& Peletier 2001,  hereafter A01).
These models have been successful in explaining some properties of spiral 
galaxies, despite their approximations. 

The current paradigm for the formation of the structure favors a 
 hierarchical clustering scenario where the structure forms by
aggregation of substructure. Hence, a galactic
object experiences the effects of collapse, merger, interaction and probably
secular evolution  in a non simple fashion, which can
also strongly depend on redshift. 
In particular, violent events (i.e.  mergers, interactions) can
have important effects on the internal properties of the
objects, such as their mass distribution (Mihos \& Hernquist 1996)
 and star formation
activity (Tissera et al. 2002,   hereafter T02).

Minor and major mergers of disc systems with satellites have 
been extensively studied.  However, most of these works do not
consider the presence of the stellar bulges. The relevance
of this component in the overall evolution of a disc system during
a merger
has been firstly pointed out by Mihos \& Hernquist (1994) in a
study of pre-prepared mergers. 
Recently, A01 studied the
effects of mergers on the structural parameters of bulges of disc-like
systems by assuming that initially bulges form with an
exponential profile. These authors found that
collisionless mergers produce a migration to higher $n$
and that this effect is proportional to the relative mass of the colliding
systems. The simulations used in this analysis were non-cosmological
pre-prepared ones where on going star formation and gas dynamics were
not included. Recently,
T02 show for the first time in 
cosmological SPH simulations how the central mass concentration
in disc-like systems can grow by collapse, merger and secular evolution.

These numerical results suggest that a
 key point in the formation of both bulges and discs in
hierarchical clustering scenarios seems to be
mergers. Mergers have been traditionally suggested as a possible
 mechanism
 to
drive the Hubble sequence since they can
 modify the dynamical and astrophysical properties of
galactic objects. Hence, the question would be how, in 
hierarchical scenarios, mergers work to shape disc-like objects
and if the properties of these objects resemble those of  current spirals.
In this paper we will focus on the analysis of the 
mass distributions of disc-like systems and how they 
are modified during mergers, paying special attention
to the comparative study of the effects of secular evolution
triggered by tidal fields
and of the actual collision of the baryonic clumps.
In order to assess the effects of mergers on the mass distributions we 
 perform a bulge-disc decomposition of
the projected surface density of the systems at different
stages of the merger process. 
We  then study how
such different structural parameters, including the
shape parameter defined by S\'ersic (1968), evolve
during the orbital decay phase and fusion of the satellite.

Our simulations include the effects of gravitation, hydrodynamics,
cooling and star formation in a cosmological framework. Hence, the set
of mergers that we analyse are given by the particular evolutionary
history of each galactic object.
This is a crucial point since the merger parameters and
physical characteristics of the colliding objects are not ad-hoc
choices but result from the consistent formation of the structure
in a hierarchical scenario. The drawback of our approach is a lower numerical
resolution compared to those used in studies of pre-prepared
mergers. Assessment of possible numerical problems are
discussed throughout the paper.

In section 2 we present the analysis of the simulations. In section
3 we discuss the results. Section 4 summarizes the conclusions.

\section{Models and Analysis}

 We run cosmological SPH simulations which include hydrodynamics and
star formation (Tissera, Lambas \& Abadi 1997).
 The simulated boxes represent typical regions of
a standard Cold Dark Matter (CDM) universe of 5 $h^{-1}$ Mpc side  and 
 $64^3$ particles ($\Omega =1$, $\Lambda =0$, $h=0.5$). 
We assume a baryonic density parameter
of $\Omega_{\rm b}=0.10$. 
All baryonic and dark matter particles have the same
mass, $M_{\rm p}=2.6\times 10^{8} {\rm M}_{\odot}$.  We have used a gravitational softening of 
$3.0$ kpc and a minimum hydrodynamical smoothing length
of $1.5$ kpc. 
We performed three simulations with different realizations of 
the power spectrum (cluster normalized)
and the same cosmological and star formation
parameters. The simulated  volume is the result of a compromise between
the need to have a well-represented galaxy sample and
enough numerical resolution to study the astrophysical properties of the
simulated galaxies.  We are confident that since 
we  focus our analysis on small scale processes such
as mergers and interactions, scale fluctuations of the order of 
the box size  will have
no significant effect on such  local processes. 
Note also that we are interested in the effects that mergers have on
the mass distributions of the galactic objects
 analysing them as individual events and
not in connection with their evolution or environment.

The star formation algorithm used in these models is   based
on the Schmidt law (see Tissera et al. (1997) for details).
 Cold gas particles are eligible to form
stars if they are denser than a certain critical value and
satisfy the Jeans instability criterium (Navarro \& White 1994).
Gas particles are checked to satisfy these conditions at all
time-steps of integration. Hence, as the gas cools down and
is gathered at the centre of dark matter haloes, it is
gradually transformed into stars according to the particular
history of evolution of each galactic object. Only one free
parameter has to be fixed: the star formation efficiency, which
links the dynamical time of the gas cloud represented by a particle
with its star formation timescale.
 We have used the value adopted
by S01 since it is adequate to reproduce disc-like structures with 
observational counterparts at $z=0$. These authors used a low 
star formation efficiency which 
allows the formation of a stellar bulge that  assures the 
axisymmetrical  character of the potential well but  
without exhausting the gas reservoir (see also DT98).
As a consequence, disc-like structures can be formed, although
they remain mainly gaseous. Conversely, observed discs are mostly stellar,
 but they
inherit the structural and dynamical properties of the gaseous discs
out of which disc stars are formed. Hence, for the sake of
comparison, the values of the parameters describing those properties
can be reliably estimated from  the simulated gaseous discs.
In this paper, we will determine and use these parameters to study the effects
of mergers on the mass distribution of the simulated galactic objects.

At $z=0$, 
we  identify galactic objects at their virial radius,   analysing
only those  with more than 4000 dark matter particles within their
virial radius in order to diminish numerical resolution problems.
 Within each galactic object, a main baryonic
clump is individualized which will be, hereafter,  called the galaxy-like
object (GLO).

The selected GLOs have   very 
well-resolved dark matter haloes 
 which provide adequate  potential wells
 for baryons to collapse in. This fact assures a reliable description of
the gas density profiles (see Steinmetz \& White 1997), which allows us
to follow the star formation history of the GLOs (see Tissera 2000).
With this strong restriction on the minimum number of particles, the
final GLO sample at $z=0$ is made up of 12 GLOs with virial velocities in 
the range 140-180 km$\cdot$s$^{-1}$.

\begin{figure}
\includegraphics[width=84mm]{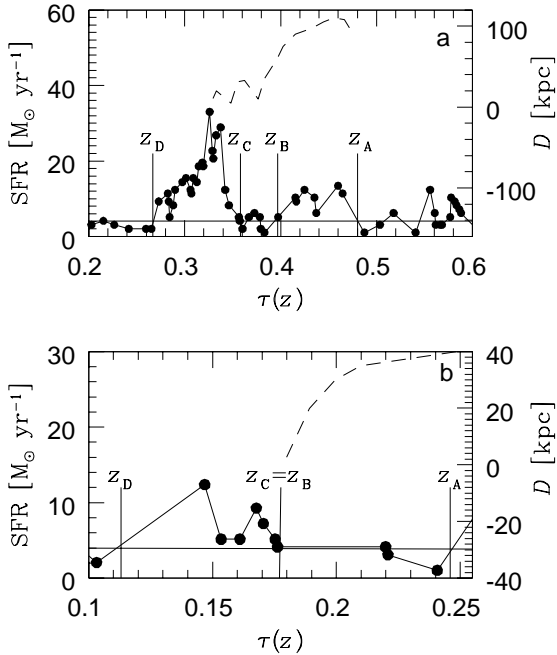}
\caption{Star formation rate (SFR) for  typical galaxy-like objects
during a double starburst (a) and a single one (b) as a 
function of look-back time ($\tau(z)$). The solid lines represent the ambient 
star formation components and the dashed lines depict the distance 
between the mass centers of the
progenitors and satellites ($D$).  
The four redshifts of interest during a merger event have been also 
plotted.}
\label{SFR}
\end{figure}

We follow the evolution of the selected GLOs with look-back time
($\tau(z)=1-(1+z)^{-3/2}$) identifying mergers and starbursts (SBs).
We then construct their mergers trees and star formation rate (SFR)
 histories (because of the restrictive SF efficiency used, these
SFR histories are mainly those of the bulge components).
 During a merger event,  the progenitor
object is chosen  as the more massive baryonic clump within this merger tree,
while the minor colliding baryonic
clump is referred to as the satellite.
Following T02 we  study only  
those mergers that are directly linked to stabursts in the SF history
of each GLO. 
These authors found that during the orbital decay phase of some merger events,
early gas inflows which trigger star formation can develop.
In this case, if the gas reservoir of the GLOs is not exhausted during these
 starbursts, second bursts are induced at the actual physical contact
of the baryonic clumps. The induction of  early gas inflows
could be directly linked to the properties of the total potential wells
of the systems in agreement with previous results reported by
Barnes \& Hernquist (1996) and  Mihos \& Hernquist (1996) in pre-prepared
simulations. When no gas inflows are triggered during the 
orbital decay phase (ODP), only one
SB is detected when the two baryonic clumps collide. These events will
be called single SBs (SSBs), while those where two SBs are detected
will be called double SBs (DSBs).
We studied a total of 18 merger events with merger parameters settled by
the cosmological model. Among them, 11 are classified as DSBs and 7 
as SSBs.
Fig. \ref{SFR}  shows an example of both types of SBs.
We have also plotted the distance between the centres of mass of
the progenitor and the satellite since the redshift they share the same 
dark matter halo.
The merger event will be determined by four redshifts of reference,
$z_{\rm A}$: the beginning of the first bursts in DSBs and the ODP
in SSBs, $z_{\rm B}$: the end of both the first bursts in DSBs and the 
ODP in SSBs, $z_{\rm C}$: the beginning of the second SBs in 
DSBs and the only  bursts in SSBs 
(in this case $z_{\rm B }= z_{\rm C}$), and $z_{\rm D}$: the end of
the second bursts in DSBs and of the only one in SSBs. 
These redshifts have been depicted in Fig. \ref{SFR}.
Note that the ODP, defined  as the time since two
objects share the same dark matter halo to the fusion of their main
baryonic cores, is determined by $z_{\rm A}$ and $z_{\rm C}$. The  $z_{\rm B}$ redshift is
used in the case of DSBs since it establishes the end of the first SBs.

In order to quantify the changes in the baryonic mass distributions, we
carried out a bulge-disc decomposition
of the projected mass surface density of the GLOs at $z=0$ and
of the progenitors at each of the redshifts that define
a merger event.
 We assumed an exponential profile for the disc component and
the S\'ersic law for the bulge central mass concentration:
\begin{equation}
 \Sigma_{\rm d} =
\Sigma_{\rm d}^{\rm o}{\rm exp}(-(r/r_{\rm d}))
\end{equation}
\begin{equation}
 \Sigma_{\rm b} =
\Sigma_{\rm b}^{\rm o}{\rm exp}(-(r/r_{\rm b})^{1/n})
\end{equation}
 where $r_{\rm d}$ and $r_{\rm b}$ are the disc and bulge scalelengths,
$\Sigma_{\rm d}^{\rm o}$ and $\Sigma_{\rm b}^{\rm o}$
the corresponding central mass surface densities and
 $n$  the bulge shape parameter.

For each GLO at $z=0$  and its progenitor objects,
 we estimated the total angular momentum ($j$)
and projected the mass distribution onto the perpendicular plane
defined by $j$. We
then integrated the projected  baryonic mass in concentric cylinders of
radius $r$ since, as
it has been proved by S01,  it is numerically more convenient
to fit the integrated mass than the surface density.  
In order to  diminish the effects of numerical resolution, the fits were
made from $r=1.5$ kpc  
 to $r\approx 30$ kpc, except in those cases where 
 the mass distributions of the progenitors were strongly perturbed
by the entrance of the satellites.

The total integrated 
baryonic mass of the GLOs ($M_{\rm GLO}$) is
estimated, on one hand, as the baryonic mass within $r=30$ kpc, and
will be associated with the luminous galaxy. On the other hand, $M_{\rm GLO}$
is the result of the disc ($M_d$) and bulge ($M_b$) mass contributions:
\begin{equation}
M_{\rm GLO}= M_{\rm d} + M_{\rm b}
\end{equation}
\begin{equation}
M_{\rm d}= 2\pi \Sigma_{\rm d}^{\rm o} r_{\rm d}^{2}
\end{equation}
\begin{equation}
M_{\rm b}= 2\pi \Sigma_{\rm b}^{\rm o}r_{\rm b}^{2} n \Gamma(2n)
\end{equation}
where  $\Gamma$ is the complete gamma function.
Since $M_{\rm GLO}$ is known,
 there are four  free parameters 
to be fitted: $r_{\rm d}$, $r_{\rm b}$, $\Sigma_{\rm b}^{\rm o}$  (or
$\Sigma_{\rm d}^{\rm o}$) and $n$.
From these fitted parameters we
analyse possible correlations with the dynamical and astrophysical properties
of the GLOs and their dark matter haloes. In particular, we
will study how the structural parameters change during merger events.
 
Note that we work with masses instead of luminosities, hence, mass-to-light
ratios for the bulge and disc  components will
have to be adopted in order to be able to compare our results 
with observations.
To this end
we use different observational results  from APB95 
(30 S0-Sbc galaxies in the $K$-band),
CdJB96 (173 spiral Sb-Sc galaxies in
deep $R$-band), Moriondo, Giovanelli \& Haynes  (1999,  hereafter MGH99, 
25 Sa-Scd spiral galaxies
in the $K$-band), KWK00
(27  early-type disc galaxies in the $K$-band) and
MCH02 (47 late-type spiral
galaxies in the $R$-band). APB95, KWK00 and MCH02
used bulge-disc decompositions of the luminosity surface density profiles,
allowing the bulge shape parameter to vary.
CdJB96 used double exponential fittings ($n=1$).
Finally, MGH99 made  fittings with different $n$ values finding that most 
of their bulges were better described by $n=1$ (except for two with
$n=2$).
Note that all these observations correspond to galaxies at $z \approx 0$.

\section{Results}
 
\subsection{GLOs at $z=0$}

As we mentioned before, S01 showed that disc-like objects formed
in numerical simulations where stellar bulges were allowed to form have
structural parameters and specific angular momentum content
similar to those of observed spiral galaxies.
For the sake of completeness and to show that these findings are  valid
for our GLOs, we will first discuss their properties at $z=0$.
The structural parameters and the 
circular  velocity ($V^{2}= G M(r)/r$) at $r=2.2 r_{\rm d}$
allow the  classification of  our objects as
intermediate spirals 
(100  km$\cdot$s$^{-1} \leq V_{2.2} \leq$ 180 km$\cdot$s$^{-1}$; 4 GLOs),
 bright spirals  ($V_{2.2}>$ 180 km$\cdot$s$^{-1}$ and 
$r_{\rm d} > 5.25$ kpc; 2 GLOs)
and compact bright spirals ($V_{2.2} >$ 180 km$\cdot$s$^{-1}$
  and $r_{\rm d} < 5.25$ kpc; 6 GLOs).
This is the result
of our strong condition on the mass of GLOs in order
to work with those better resolved.

\begin{figure}
\includegraphics[width=84mm]{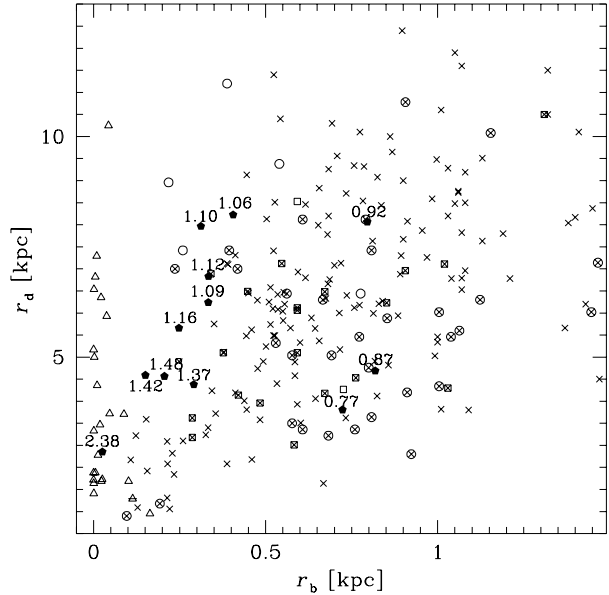}
\caption{Disc scalelength ($r_{\rm d}$) as a function 
of bulge scalelength ($r_{\rm b}$)
for simulated objects at $z=0$ (filled pentagons). Labels with 
the shape parameters
($n$) of the simulated bulges are shown. 
Observational data
from CdJB96 (crosses), MGH99 ($n=1$ crossed squares, $n=2$: open squares),
KWK00 ($n\le 1.5$: dashed triangles, $n>1.5$: open triangles) and 
 MCH02 ($n\le 1.5$: crossed circles, $n>1.5$: open circles) 
 are also  included. }
\label{rbrd0}
\end{figure}

In Fig. \ref{rbrd0} we show the relation between the disc and bulge 
scalelengths. Each GLO has a label that specifies its $n$ shape parameter.
 We include the observational data from CdJB96,
  MGH99, KWK00  and MCH02.
Note that for increasing $n$ parameters, the scalelengths tend to move
to the upper envelope of the  distribution. This effect is also seen
in the observed scalelength distribution.
We have adopted  $n=1.5$ as a general cut-off value to segregate 
the samples between those with exponential and non-exponential profiles
for the bulge components.

\begin{figure*}
\includegraphics[height=69mm]{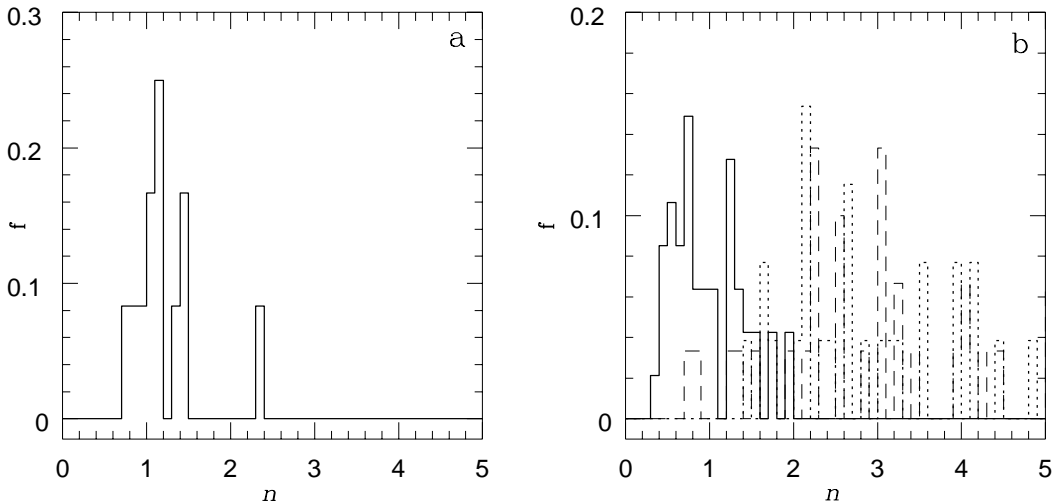}
\caption{Histogram of the shape parameter of the bulges for simulated
objects at $z=0$ (a) and observations from APB95 (dashed line),
KWK00 (dotted line) and MCH02 (solid line), normalized to the total
number of galaxies in each sample.}
\label{hist0}
\end{figure*}

In Fig. \ref{hist0} we show the normalized
histrogram of $n$ for GLOs at $z=0$ (a),
and  the corresponding distribution for the observations of 
APB95, KWK00 and MCH02 (b), normalized to the total number of galaxies
in each sample. 
The combined observational sample covers a morphological
range from Sa to Sc galaxies, but this is not a consistent sample.
We show it in order to broadly compare the range of $n$ values
obtained from the simulations with observations.
The average $n$ values  over the observed subsamples are
$<n_{\rm APB95}>= 2.97 \pm 1.29$, $<n_{\rm KWK00}>= 2.88 \pm 0.94$
and
$<n_{\rm MCH02}>= 1.02 \pm 0.43$ while the simulated sample has  
$<n_{\rm sim}>=1.23 \pm 0.40$. The simulated distribution is consistent
with the observations of MCH02 at one $\sigma$-level, which suggests
that our objects are more similar to late-type galaxies.
However,  $<n_{\rm sim}>$ value also agrees at $3$ 
$\sigma$-level with the corresponding
 mean values of the early-type samples.

\begin{figure}
\includegraphics[width=84mm]{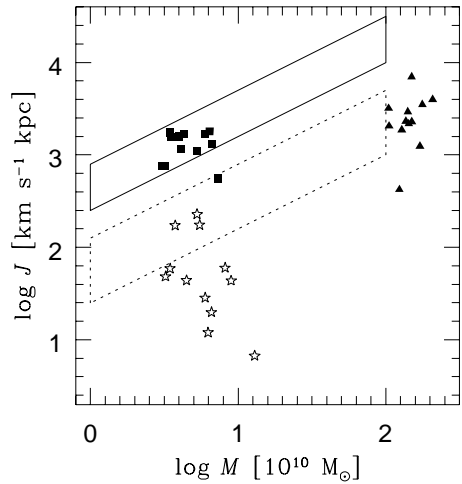}
\caption{Specific angular momentum ($J$) for the gaseous
discs at $3.2 r_{\rm d}$ (filled squares),
the stellar bulges at $r_{\rm eff}$
(open stars) and dark matter haloes at the virial radius (filled triangles),
as a function of mass ($M$).
The solid (dotted) box depicts the observational area for
spirals (ellipticals)
from Fall (1983).}
\label{mj}
\end{figure}

A key point in the formation of discs is the conservation of 
the specific angular momentum (FE80). Our SF
algorithm has been implemented in such a way that it has allowed
the formation of stellar bulges but without exhausting the gas reservoir
(DT98).
Hence, gaseous discs have been able to regenerate after mergers.
In order to compare the angular momentum ($j$) of the
three mass components: bulge, disc and dark matter halo, with observations, we
plot  
in Fig. \ref{mj}  their specific angular momentum ($J$=$j/M$) 
content as a function of  mass ($M$).
The mass and angular momentum of bulges have been estimated
with stars within $r< r_{\rm eff}$, while those of the
gaseous discs have been calculated
within $r< 3.2\  r_{\rm d}$. These criteria have
been adopted following those used by observers. The mass and
angular momentum of the dark matter haloes have been estimated
at the virial radius. 
In this figure, we have also included two boxes that depict the observational region 
covered by spirals and ellipticals as given by Fall (1983).
From this plot we can see that the disc components and the dark matter
haloes have comparable specific angular momentum contents as predicted
by the standard disc formation model of FE80.
Conversely,  stellar bulges have been formed from material that
has lost most of its angular momentum.

From these results we conclude that the simulated GLOs have
 structural and dynamical parameters
that statistically resemble those of current normal spirals at $z=0$.
 In the following section we will study the role played by
mergers in the determination of these properties and
try to understand the origin of correlations among
the structural parameters such as $B/D$ versus $n$.

\subsection{Analysis of progenitors during merger events}

We carried out the  bulge-disc (B+D)
 decomposition of the progenitor objects of the GLOs
analised in the previous section  during
merger events at the four redshifts of interest:
$z_{\rm A}$, $z_{\rm B}$, $z_{\rm C}$ and $z_{\rm D}$.
In this section we will study how their structural parameters
change during these processes.

\subsubsection{The shape parameter}

\begin{figure*}
\includegraphics[height=130mm]{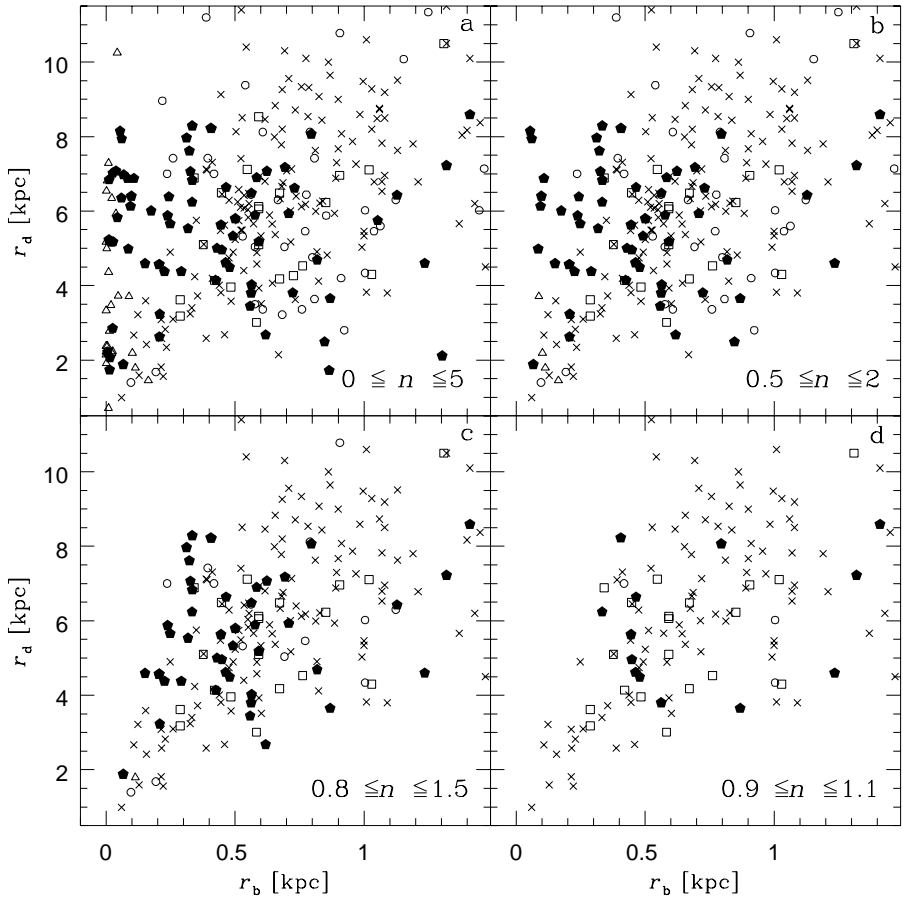}
\caption{Disc scalelength ($r_{\rm d}$) 
as a function of bulge scalelength ($r_{\rm b}$)
for  simulated objects at $z=0$ and progenitors (filled pentagons) 
for which bulges
are described by $0\le n\le 5$ (a), $0.5\le n\le 2$ (b), $0.8\le n\le 1.5$ (c) and 
$0.9\le n\le 1.1$
(d). Observational data
from CdJB96 (crosses), MGH99 (open squares),
KWK00 (open  triangles) and 
 MCH02 (open circles) for the same shape parameter
intervals are also shown.}
\label{nrest}
\end{figure*}

In Fig. \ref{nrest}(a)  we show the $r_{\rm d}$ versus $r_{\rm b}$
relation for all
progenitors during merger events and at $z=0$.
We have also included the observational 
data from a B+D decomposition from  MGH99, KWK00 and MCH02, and
from a double exponential decomposition by CdJB96.
The combined observed sample of spirals shows
a certain correlation between $r_{\rm d}$ and $r_{\rm b}$,
although some of them  have, for a given disc scalelength,
a smaller bulge one than that expected from this relation.
  We also note that the scalelengths of
our simulated disc structures show a similar behaviour. In order to 
individualize which observed and simulated objects belong to each
of these two distributions of scalelengths,
we segregate them according to their shape parameters as shown from 
Fig. \ref{nrest}(b) to Fig. \ref{nrest}(d). 
It is clear that as galaxies and simulated objects with large shape parameters
are taken out, the relation for both,  observed and simulated 
scalelengths, gets better defined. 

We also note that the $r_{\rm b}$ versus $r_{\rm d}$ relation 
for GLOs and galaxies with bulges with $n>1.5$ 
 shows no trend at all.
It is encouraging that the  observed and simulated disc structures
show  scalelength distributions with similar patterns,
 for both $n\approx 1$ or $n\ne 1$ bulge profiles. 
In the case of exponential bulges ($0.9\le n\le 1.1$), the correlation
is well defined with $<r_{\rm b}/r_{\rm d}>= 0.14 \pm 0.05$ and
 $<r_{\rm b}/r_{\rm d}>= 0.13 \pm 0.06$ for observations and simulations,
respectively.
Table \ref{porcent} summarizes the mean values of $r_{\rm b}$, 
$r_{\rm d}$ and $n$
for GLOs at the four $z$ of interest.

\begin{table}
\center
\caption
{Mean values of bulge scalelength ($r_{\rm b}$), disc scalelength 
($r_{\rm d}$)
and bulge shape parameter ($<n>$)
for single (S) and double (D) SBs for the redshifts of
interest.}
\vspace{0.4cm}
\begin{tabular}{|c c| c| c| c | }\hline
$z_i$ & & $r_{\rm b}$ [kpc] & $r_{\rm d}$ [kpc]& $<n>$  \\\hline
$z_{\rm A}$ & S & 0.46$\pm$0.17 & 6.2$\pm$0.7 & 1.5$\pm$0.2\\
      & D & 0.31$\pm$0.11 & 4.9$\pm$0.6 & 1.8$\pm$0.2 \\\hline
$z_{\rm B}$ & S & 0.57$\pm$0.16 & 5.6$\pm$0.7 & 1.3$\pm$0.2 \\
      & D & 0.37$\pm$0.16 & 5.9$\pm$0.4 & 1.6$\pm$0.2 \\\hline
$z_{\rm C}$ & S & 0.57$\pm$0.16 & 5.6$\pm$0.7 & 1.3$\pm$0.2 \\
      & D & 0.53$\pm$0.12 & 5.0$\pm$0.4 & 1.2$\pm$0.1  \\\hline
$z_{\rm D}$ & S & 0.52$\pm$0.12 & 5.0$\pm$0.7 & 1.4$\pm$0.3 \\
      & D & 0.37$\pm$0.08 & 5.4$\pm$0.6 & 1.4$\pm$0.2 \\\hline

\end{tabular}
\\
{\small
Note: The errors correspond to the bootstrap method, $\sigma_{\rm bt}$. }
\label{porcent}
\end{table}

\begin{figure}
\includegraphics[width=84mm]{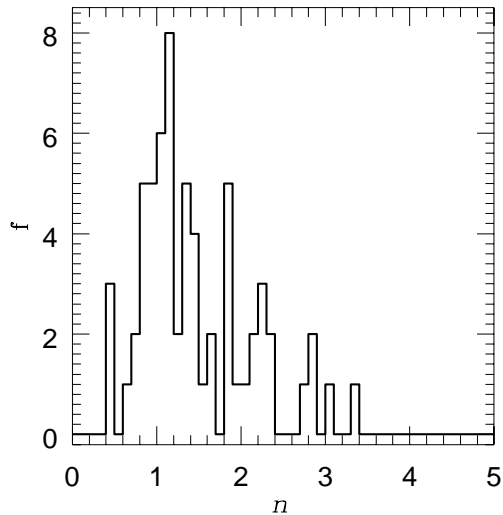}
\caption{Total distribution of bulge shape parameters for the progenitors
during merger events.}
\label{hist}
\end{figure}

Similary to Fig. \ref{hist0}(a) we display in Fig. \ref{hist} the histogram
of shape parameters for all progenitors at all studied redshifts. 
Hence 
this distribution  gathers the information on the $n$ parameters of different progenitors 
at different stages of evolution. From the comparison with that
shown in Fig. \ref{hist0} corresponding to  the bulges of the
final structures at $z=0$, we deduce that in the past of current GLOs,
their progenitors went through stages where their
bulges had  $n>1.5$ parameters ($\sim 35$ per cent of the
total sample). Meanwhile,
 GLOs at $z=0$ show a trend
to have $n\approx 1$, with less than $10$ per cent being described by
 $n>1.5$.

In order to assess the effects of mergers on the shape parameters
of the objects and to look for possible differences between the
effects of secular evolution and fusions, we calculate the
percentages corresponding to  the  increase and decrease of the
shape parameter during the intervals  $z_{\rm A}$-$z_{\rm C}$, $z_{\rm C}$-$z_{\rm D}$ and
 $z_{\rm A}$-$z_{\rm D}$ for SSBs and DSBs.
The results are shown in Table \ref{nchange}. 
Note that we are taking the interval  $z_{\rm A}$-$z_{\rm C}$ as the
total period corresponding to the ODP.
For both single and double SBs, we find that bulges tend to decrease their
$n$ parameters during the ODP, and that this effect  is more important for DSBs.
In contrast, the actual fusion of the clumps ($z_{\rm C}$-$z_{\rm D}$) 
tends to make the $n$ value of the bulges larger. 

\begin{table}
\center
\caption{Percentage of objects which increase ($\uparrow n$)/decrease
($\downarrow n$) their
shape parameter during the ODP ($z_{\rm A}$-$z_{\rm C}$), the fusion of
the baryonic clumps ($z_{\rm C}$-$z_{\rm D}$) and the whole merger process
($z_{\rm C}$-$z_{\rm D}$), for single (S) and  double (D) starbursts.}
\begin{tabular}{|c c| c| c|c|}\hline
 && $z_{\rm A}$-$z_{\rm C}$ [\%] &$z_{\rm C}$-$z_{\rm D}$ [\%] 
&$z_{\rm A}$-$z_{\rm D}$ [\%] \\\hline
S& $\uparrow n$& 42.9 & 57.1 & 57.1 \\
& $\downarrow n$& 57.1 & 42.9 & 42.9 \\\hline
D& $\uparrow n$& 18.2 & 63.6 & 45.5 \\
& $\downarrow n$& 81.8 & 36.4 & 54.5 \\\hline
\end{tabular}
\label{nchange}
\end{table}

Therefore we find that the physical encounter of the baryonic cores tends to 
produce bulges with larger shape parameters (i.e.  more concentrated),
and $n\approx 1$ profiles tend to be
formed by secular evolution.
These results  together with the fact that at $z=0$ most of
GLOs show $n\le 1.5$, regardless of their past history, suggest
that successive mergers have driven the bulges from 
exponential to more concentrated profiles, to come back to 
exponential ones, if there is available gas in the systems. 

We think that the possibility of
having secular evolution and core fusion associated to
 a merger  helps to drive this cycle in the shape parameter.  
Note that the collisionless merger of stellar discs
will not produce the same distributions. As a matter of fact,
in order to detect this {\it morphological loop}, simulations should include
gas dynamics and star formation since, otherwise, bulges can not
be re-shaped back to $n \approx 1$ (A01).

\begin{figure*}
\includegraphics[width=130mm]{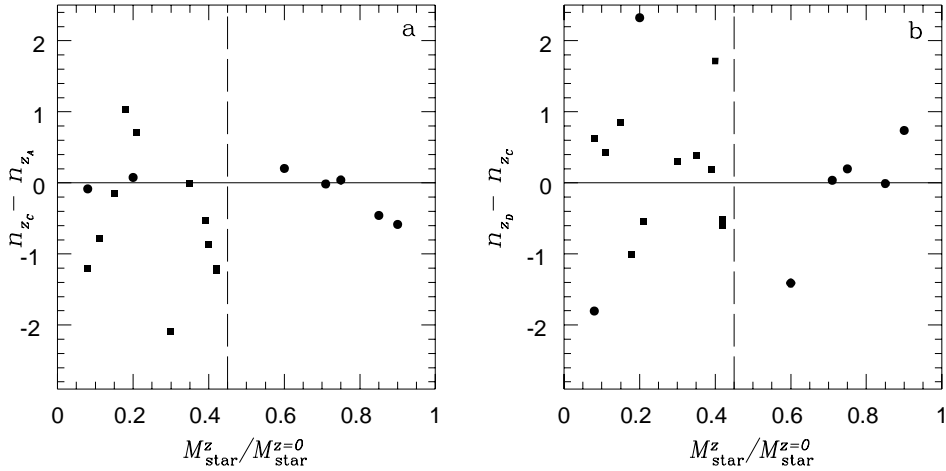}
\caption{Variations in the shape parameter of  the
progenitors  
during the orbital decay phase (a) and during the fusion of the baryonic
clumps (b) in single (filled circles) and double (filled squares) SBs,
 as a function
of the fraction of stars already formed in the progenitors,
$M^{\rm z}_{\rm star}/M^{\rm z=0}_{\rm star}$ when the mergers start.
$M^{\rm z}_{\rm star}/M^{\rm z=0}_{\rm star}=0.45$ (dashed lines) 
has been taken to be an indicator of the presence of a  well-formed
stellar bulge after T02.}
\label{nmstar}
\end{figure*}

Fig. \ref{nmstar} shows the variations of  the $n$ parameter during
the ODP (a) and those produced during  the fusion  of the baryonic
 clumps (b)  as
a function of $M_{\rm star}^z/M_{\rm star}^{z=0}$, where
 $M_{\rm star}^z$ is the stellar mass formed in the progenitors
at $z$ and $M_{\rm star}^{z=0}$
is the total stellar mass of the GLOs at $z=0$. This ratio
is an estimate of the presence of a stellar bulge since in these
simulations the stars
form preferentially in the dense regions.
T02  found that the properties of the potential
well can be directly linked to the stability of a system,
in the sense that the shallower the potential
well, the more susceptible the GLO to experience early gas inflows
(i.e.  secular evolution). The formation of
stellar bulges has been found to inhibit such early gas inflows. This picture 
 is in agreement with our results, where
the major changes in the shape parameter during a merger  event correspond
to the GLOs that experience DSBs ($M_{\rm star}^z/M_{\rm star}^{z=0}\le 
0.45$), which are less stable objects
and can be strongly perturbed by
the  collision with a satellite. In contrast, more
stable systems ($M_{\rm star}^z/M_{\rm star}^{z=0}>0.45$)
would not experience such important perturbations 
in its mass concentration during the ODP (Table \ref{nchange}).

We can also see from Fig. \ref{nmstar}(a) that the most of the
 changes in $n$ are negative and can be
linked to smaller  $M_{\rm star}^z/M_{\rm star}^{z=0}$ values
($M_{\rm star}^z/M_{\rm star}^{z=0}\leq 0.45$).
This could be  related to the fact that stellar bulges
can provide stability to the systems, making them less vulnerable
to the influence of a  satellite, principally during the ODP.
From Fig. \ref{nmstar}(b) we see that the fusion of the baryonic cores
produce more positive changes in the $n$ parameters and hence,
a trend to increase it. Again, the larger variations are related
to smaller bulges (i.e.  smaller  $M_{\rm star}^z/M_{\rm star}^{z=0}$).

We also analysed the possibility that the changes in the
$n$ value during mergers are linked to the relative  mass
of the colliding systems, finding
no correlation in contrast to the results of A01.
However, these authors only considered collisionless mergers while
in all our mergers dissipation plays an important role.

\begin{figure}
\includegraphics[width=84mm]{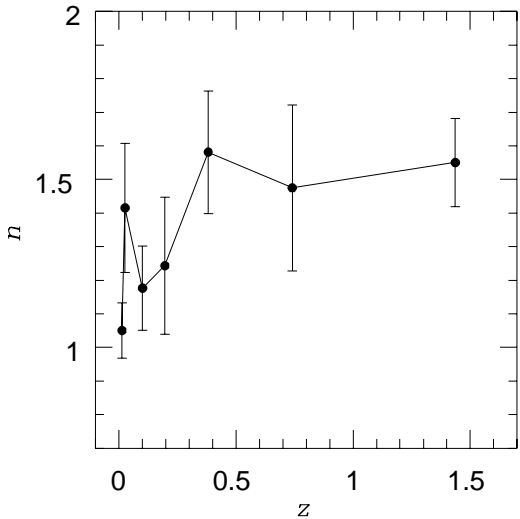}
\caption{Average $n$ parameters of the progenitors at different
$z$ intervals and of the GLOs at $z=0$. The error bars correspond
to the bootstrap method. }
\label{nz}
\end{figure}

In Fig. \ref{nz} we have plotted the averaged $n$ parameters of the
progenitors at different $z$ intervals.  We can see that, although
the bootstrap errors are high, there is a trend for the averaged $n$  to
decrease toward $n=1$ for decreasing $z$.
Nevertheless, at all redshifts we found the mean $n$ smaller than
$n=2$, indicating that, in spite of the spread of the parameters,
the majority of the bulges are best fitted with $n\la 1.5$.

\subsection{The $r_{\rm b}/r_{\rm d}$ ratio}

It has been argued by some authors (e.g., CdJB96)
that there is a restricted range of $r_{\rm b}/r_{\rm d}$ for 
spiral galaxies which has been interpreted as the result of secular
evolution of the central mass concentrations. In Fig. \ref{rbrd0}
we showed the $r_{\rm d}$ versus $r_{\rm b}$ distribution for GLOs at
$z=0$ and
for   
observations. 
We have combined the observational results from MGH99, KWK00
and MCH02 which include spiral galaxies from Sa to Sc.
For these distributions we find 
$<r_{\rm b}/r_{\rm d}>=0.07\pm 0.02$
(the boostrap error is $\sigma_{\rm bt}=0.02$) and 
$<r_{\rm b}/r_{\rm d}>=0.10\pm 0.08$ ($\sigma_{\rm bt}=0.01$)
for the simulated and observed data, respectively, showing a good
agreement. 

\begin{figure}
\includegraphics[width=84mm]{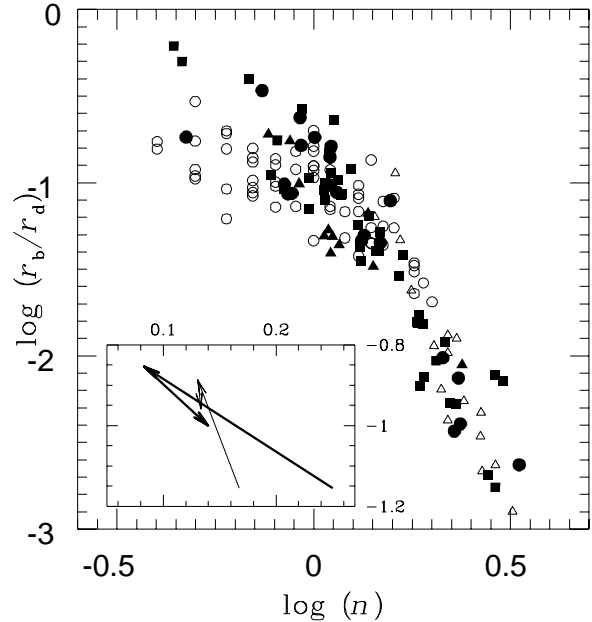}
\caption{Ratio between bulge and disc scalelengths as a function
of the bulge shape parameter  for simulated objects at
$z=0$ (filled triangles), single SBs (filled circles) and double
ones (filled squares).
Observational data
from KWK00 (open triangles) and 
 MCH02 (open circles) are also shown.
The minor box shows the changes of the plotted parameters for single
(thin line)  and double (thick line) SBs during the
orbital decay phase (first arrow) and the fusion of the baryonic clumps
(second arrow). }
\label{nrbrd}
\end{figure}

However, we have already shown in Fig. \ref{nrest} that
 the simulated and observed scalelength distributions
 show a clear trend only
for those objects with bulges with $0.9\le n\le 1.1$.
Hence, we divided the observations and simulated data
into three subsamples according to their $n$ parameters
($n<0.9$, $0.9\le n\le 1.1$ and $n>1.1$) and estimate their mean 
$r_{\rm b}/r_{\rm d}$. Table \ref{rbrd} summarizes the results.
First, we note that observational and simulated
$<r_{\rm b}/r_{\rm d}>$ in the three subsamples are in very good
agreement, supporting the hypothesis that GLOs in CDM models have 
observational counterpart. We also see that the larger 
dispersion is found for bulges with $n<0.9$ in both the observed and
simulated data, and that these subsamples show  higher
$<r_{\rm b}/r_{\rm d}>$ than the subsamples with $n \approx 1$,
suggesting that these systems systematically
have  larger $r_{\rm b}$ (for this subsample:
$<r_{\rm b}>=0.89\pm 0.10$ and $<r_{\rm d}>=4.77 \pm 0.40$). 

\begin{figure*}
\includegraphics[height=130mm]{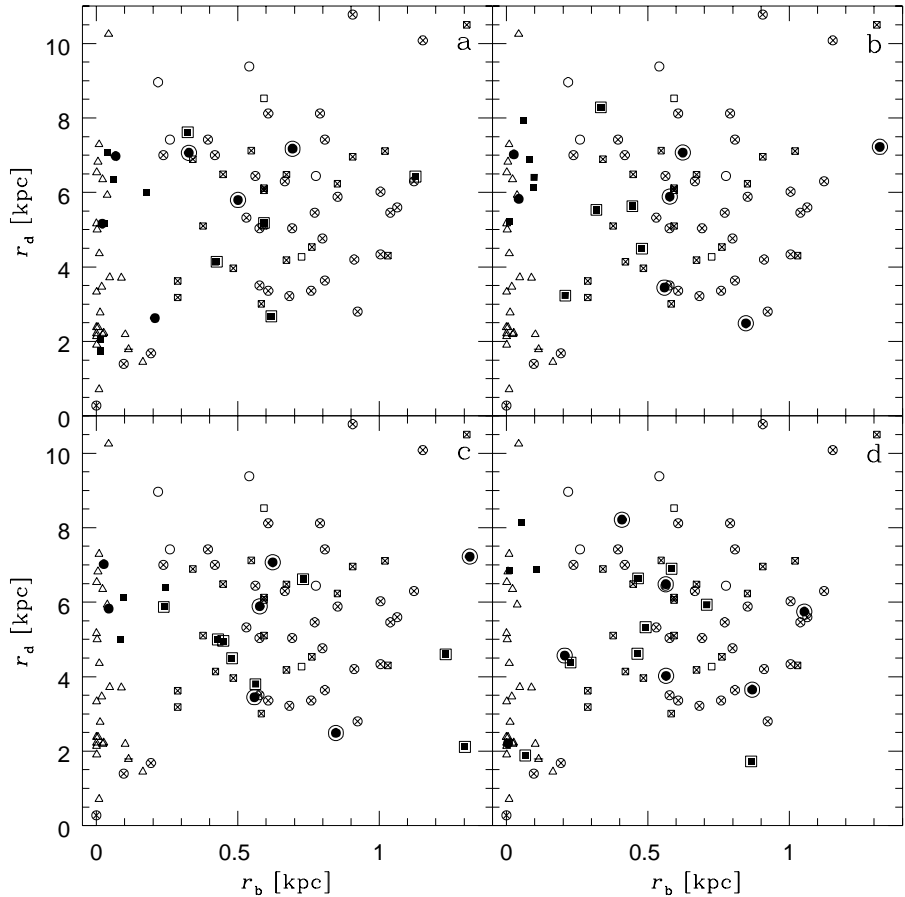}
\caption{ Distribution of bulge and disc scalelenghts for
the simulated  galactic objects at $z_{A}$ (a),  $z_{B}$ (b),
 $z_{C}$ (c) and  $z_{D}$ (d). Different symbols have
been used  to distinguish between $n\le 1.5$ (encircled symbols) and
$n>1.5$, and between double (filled squares) and single (filled
circles) starbursts.
Observational data
from MGH99 ($n=1$ crossed squares, $n=2$: open squares),
KWK00 ($n\le 1.5$: crossed-triangles, $n>1.5$: open triangles) and 
 MCH02 ($n\le 1.5$: crossed-circles, $n>1.5$: open circles) are also shown.}
\label{rbrdSD}
\end{figure*}

On the other hand, we find that systems with bulges better fitted 
by $n>1.1$ show the smallest $<r_{\rm b}/r_{\rm d}>$. Hence for 
a given disc scalelength, the larger the concentration of the bulges,
the smaller its scalelength (for this subsample:
$<r_{\rm b}>=0.23 \pm 0.03$ and $<r_{\rm d}>=5.52 \pm 0.30$).
If we assume that the $n$ parameter
correlates with morphology as it is indicated by several works
(e.g., de Jong 1996; Graham \& Prieto 1999, hereafter GP99; GdB01), 
then the scalelength ratios should correlate with $n$.
 Fig. \ref{nrbrd}  
shows  the $r_{\rm b}/r_{\rm d}$ ratio as a function of the shape
parameter for the simulated objects during mergers 
and for $z=0$. We have also included observational data from
 KWK00 and  MCH02. We find an anticorrelation between 
 these two parameters, in agreement with observations.
Assuming that the shape parameter is an indicator
of  Hubble type, this
anticorrelation indicates that late-type spirals  ($n\leq 1.5$) 
have larger  $r_{\rm b}/r_{\rm d}$ ratios than early-type spirals
($n>1.5$). We also note that the relation for the observed and simulated
ratios  seems to flatten for
$n\ll 1$. A simple extrapolation gives a  value of $r_{\rm b}/r_{\rm
d}\approx 0.40$. 

Because the shape parameter and the scalelength of the S\'ersic
law are coupled parameters, this trend of 
$r_{\rm b}/r_{\rm d}$ with $n$ could be the result of the fitting formula.
However, Trujillo, Graham \& Caon (2001, hereafter TGC01) 
argued that there is a physical
trend stronger than that resulting from the fitting formula.
Because we use the same expression for all fittings, the fact
that there is a segregation in the scalelength distribution
according to the value of $n$ suggests that this trend could have
physical basis and could be related to a difference in the
evolution of the GLOs or in the efficiency of some physical process. 
This is confirmed when $r_{\rm eff}$ is used instead of $r_{\rm b}$
for which we find  $<r_{\rm eff}/r_{\rm d}>=0.29$  ($\sigma_{\rm bt}$=0.04)
for $n< 0.9$ 
, $<r_{\rm eff}/r_{\rm d}>=0.26$  ($\sigma_{\rm bt}$=0.03)
 for $0.9\leq n\leq 1.1$, and  
$<r_{\rm eff}/r_{\rm d}>=0.30$  ($\sigma_{\rm bt}$=0.05) for $n > 1.1$ 
(the combined observational sample of KWK00 and MCH02
shows $<r_{\rm eff}/r_{\rm d}>=0.23$  ($\sigma_{\rm bt}$=0.02)
for $n< 0.9$,  $<r_{\rm eff}/r_{\rm d}>=0.28$  ($\sigma_{\rm bt}$=0.04)
 for $0.9\leq n\leq 1.1$, and  
$<r_{\rm eff}/r_{\rm d}>=0.35$  ($\sigma_{\rm bt}$=0.03) for $n > 1.1$).
We have used the approximation of
Ciotti \& Bertin (1999) to transform simulated
$r_{\rm b}$ into $r_{\rm eff}$. 

\begin{table}
\center
\caption{Mean values of the ratio between
 bulge and disc scalelengths for simulations (SIM) and
 observations (OBS) for different cases. S and D denotes single and double
SBs, respectively. The bootstrap errors are shown in brackets.}
\begin{tabular}{|c| c| c|}\hline
 & $<r_{\rm b}/r_{\rm d}>_{\rm SIM}$ &$<r_{\rm b}/r_{\rm d}>_{\rm OBS}$ \\\hline
$z\neq 0$ & 0.10$\pm$0.10 (0.01) & \\
$z=0$ & 0.07$\pm$0.02 (0.02) & $0.10\pm 0.080$ (0.008)\\\hline
$n<0.9$ &$0.24\pm 0.20$ (0.05) &$0.16\pm 0.07$ (0.01)\\
$0.9\le n\le 1.1$ &$0.13\pm 0.06$ (0.02) &$0.14\pm 0.05$ (0.02)\\
$n>1.1$ &$0.05\pm 0.05$ (0.01) &$0.03\pm 0.04$ (0.01)\\\hline
S $z_{\rm A}$ & 0.07$\pm$0.05 (0.02)& \\
S $z_{\rm D}$ & 0.11$\pm$0.08 (0.03)&\\\hline
D $z_{\rm A}$ &0.07 $\pm$0.08  (0.02)&\\
D $z_{\rm D}$ &0.10$\pm$0.10 (0.04) & \\\hline
\end{tabular}
\label{rbrd}
\end{table}

Let us now look at the mean $<r_{\rm b}/r_{\rm d}>$ during mergers.
In Table \ref{rbrd} we show the mean  $r_{\rm b}/r_{\rm d}$ at $z_{\rm A}$
and $z_{\rm D}$ for SSBs and DSBs. We find no differences between the
mean $r_{\rm b}/r_{\rm d}$ but both samples agree in increasing
the mean ratios and the dispersions at $z_{\rm D}$.
Fig. \ref{rbrdSD} shows the distributions of scalelength during
mergers. Note that, independently  of the stage of evolution
of the mergers, the distributions have similar characteristics
and are within the observed range at $z=0$.
Finally in Fig. \ref{nrbrd}, in the small box, we show how the averaged
 $r_{\rm b}/r_{\rm d}$ moves throughout  mergers
 associated with single (thin line) and
double (thick line) starbursts. It is clear that
early gas inflows during the ODP tend to decrease $n$ in both cases
but for DSBs the changes are more important, as expected from their
stability properties (see Fig. \ref{nmstar} and T02).
Conversely, the fusion of the cores moves
the distribution toward higher $<n>$.

\subsection{The $B/D$  ratio}

The luminosity $B/D$ ratio has been traditionally used as an indicator of 
morphology (e.g.,
Andredakis \& Sanders 1994).
The correlation between $B/D$ and $n$ has been 
 shown by APB95 and recently
confirmed by TGC01. Linear regression through observations
show $\partial (B/D)/{\partial n}= 1.0\pm 0.3$\   ($\sigma_{\rm
bt}=0.3$)
 for
APB95, $0.8\pm 0.5\  (\sigma_{\rm bt}=0.5)$ for KWK00 and 
$1.7\pm 0.3 \  (\sigma_{\rm bt}=0.3)$ for MCH02.
The combined sample has  $\partial (B/D)/ {\partial n}= 1.2\pm
0.1\  (\sigma_{\rm bt}=0.1)$.

Note that in the models we have masses instead of luminosities and that
the mass-to-light ratios of  the discs ($\gamma_{\rm d}$) 
and bulges ($\gamma_{\rm b}$) could be different
because of their different stellar populations.
However, their correct values and/or  possible dependence on
radius as well as redshift are still unclear.
In order to match the luminosity $B/D$ we have to rescale the mass
$B/D$ ratio by a constant factor of 20.
We address that we are using the same mass-to-light ratios
for all GLOs regardless of the redshift\footnote{Nevertheless, 
we found that very similar $\gamma_{\rm b}/\gamma_{\rm d}$ 
ratios are needed to match the observed luminosity $B/D$ for progenitors at
different $z$, although our dispersion is quite high.
Note that we have used the total baryonic masses to estimate
luminosities. In a more realistic model including energy
feedback, some fraction of the cold gas would be reheated.
Hence, the values required for our models to
match the observed luminosities could be taken as upper limits
to the mass-to-light ratios.}, disc scalelength or 
observational band.

\begin{figure*}
\includegraphics[height=130mm]{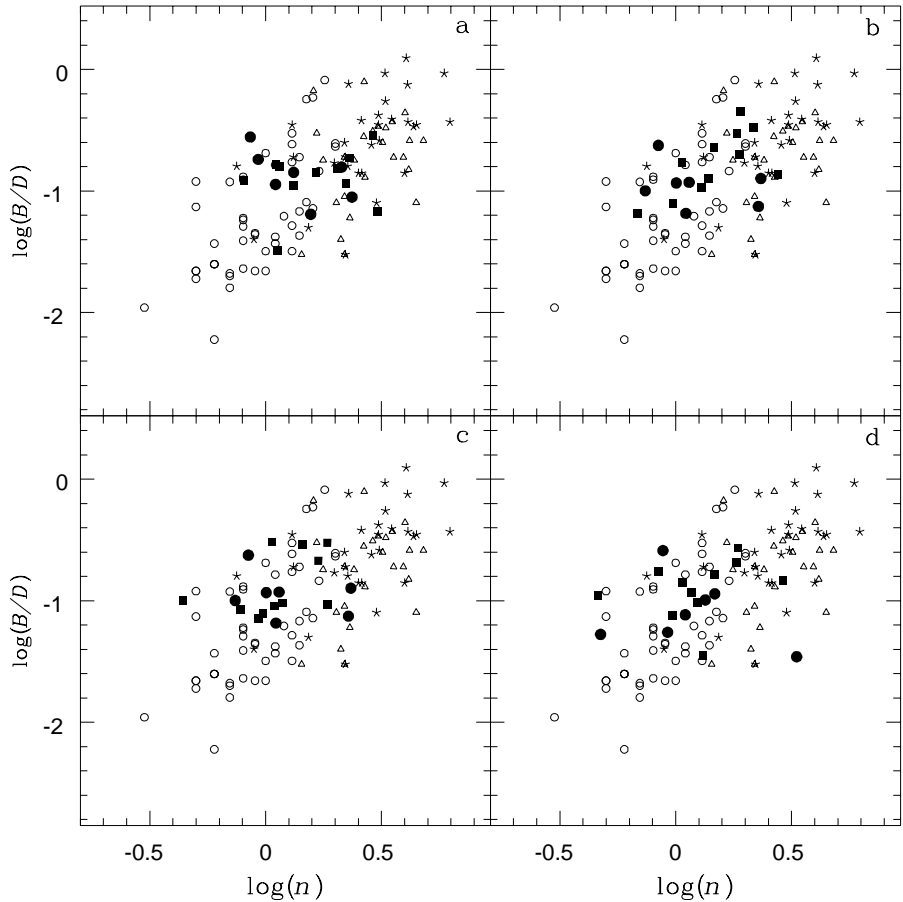}
\caption{Luminosity bulge-to-disc ratio ($B/D$) as a function of the shape
parameter of the progenitors during single (filled circles) and 
double (filled squares) SBs 
at $z_{\rm A}$ (a), $z_{\rm B}$ (b), $z_{\rm C}$ (c) and
  $z_{\rm D}$ (d).  
Observational data
from APB95 (asterisks),
KWK00 (open triangles) and 
 MCH02 (open circles) are also shown.}
\label{bdn}
\end{figure*}

In Fig. \ref{bdn} we show  the $B/D$ ratio for the simulated GLOs
during  merger events at the four redshifts of interest
and the observational data from APB95, KWK00 and MCH02.
The low statistical number restrict the use of linear regression through
the simulated data.
However,  we see that, at any time during merger events, the simulated
values are within the observational range. 

\begin{figure*}
\includegraphics[width=140mm]{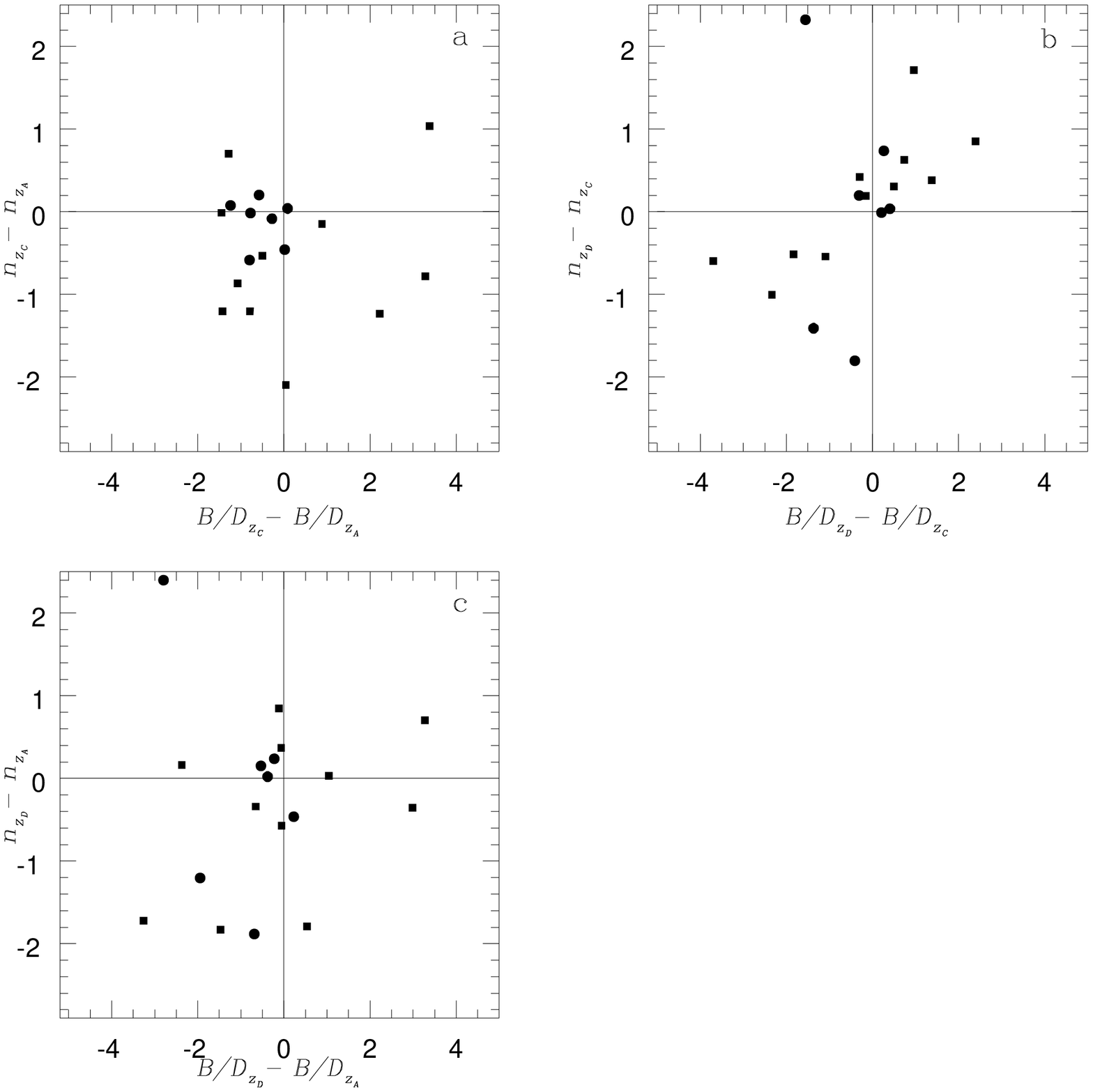}
\caption{Variations in the shape parameters of the
progenitors 
 during the orbital decay phase (a), the fusion of the baryonic clumps
(b) and the whole merger process (c) in single (filled circles)  and double (filled squares)
SBs  as a function of the 
corresponding changes in the bulge-to-disc ratio, $B/D$. }
\label{nbdchange}
\end{figure*}

In  Fig.\ref{nbdchange} we plot 
 the variations 
$n_{z_i}-n_{z_j}$ versus $B/D_{z_i}-B/D_{z_j}$  where $i$/$j$ can be
A, C or D.
From this figure we see that  GLOs
which experience gas inflows show more important changes in both $n$
 and $B/D$
during the ODP.
In the case of those  GLOs experiencing SSBs (e.g.,
no early gas inflows) there is no important change in the 
 shape parameter or the $B/D$ ratios during the ODP.
We also note that the changes in $n$ and $B/D$ do not correlate.
Conversely, during the actual fusion 
of the objects (from $z_{\rm C}$ to $z_{\rm D}$)
both types of GLOs experience large changes in  both parameters
in the expected direction: the larger the increase in the shape parameter,
the larger the corresponding increase of 
the $B/D$ ratio or vice-versa.  Fig. \ref{nbdchange}(c) shows the overall
changes over the merger events. 
This distribution has higher dispersion than that obtained during
the fusions because it also combines  the effects of secular evolution.
However, the overall effects of mergers is to settle the correlation
between $B/D$ and $n$ so that more concentrated bulges (higher shape
parameters) have the larger $B/D$ ratios. However, we note
that, although the  range of values can be matched,
the simulated $B/D$ do not show the same correlation signal ($r=0.20$)
than the observations ($r=0.55$).

\section{Conclusions}
   
We have studied the properties of the baryon distributions
in galaxy-like objects focusing on the effects of mergers.
We resort to observations of spiral galaxies of different morphology
to constrain our findings.

We found that on average, galactic objects formed in hierarchical
clustering scenarios reproduce the angular momentum 
and structural parameter distributions of spiral galaxies, if
a stellar bulge is allowed to form and early gas depletion is avoided.
We have succeeded in these two aspects but at the expense of
inhibiting star formation on discs. A consistent treatment of energy
feedback may help to remove this caveat.

These simulations have allowed us to study how mergers change the
distribution of baryons by analysing the evolution of their
structural parameters.
We found that, on average,
galactic objects tend to  have nearly exponential bulges
 at all redshift. 
However, note that these simulations produce
 bulges with shape parameters in the
range $0.5 - 4$.

For those systems with $n\approx 1 $ bulges we found a
correlation
among their bulge and disc scalelengths in very good agreement with
observations. However, for $n>1$ the scalelength distribution  is
disorder and displaced to smaller bulge scalelenghts. Observations
show the same behaviour. The opposite distribution is found for 
systems with $n<1$ bulges  which have larger $r_{\rm b}/r_{\rm d}$ values 
and larger dispersions.
We found the disc scalelengths to be approximately
 independent of $n$ parameters,
so that this correlation implies that more concentrated objects
have smaller  bulge scalelengths (or large effective radius).
We also found a maximum  $r_{\rm b}/r_{\rm d}$  of $\approx 
0.40$ for $n\rightarrow0$. Higher numerical resolution simulations are needed
to study the formation of such low mass surface profiles.

In order to test our results for low resolution problems in the
determination of the structural parameters of the galaxy-like
objects, we followed Steinmetz \& M\"uller (1994) and used
the bootstrap technique to estimate the effects of low particle 
number statistics
within the objects. We calculated an accuracy better than 25 per cent
for the scalelengths and shape parameters.

We found that gas inflows during the orbital decay phase tend
to produce important changes in the mass distributions generating
$n\approx 1$ profiles, while the fusions of the baryonic
cores tend to increase the $n$ parameter.
As a consequence, a morphological loop can be driven by mergers
which might be responsible of triggering secular evolution
as well as of the baryonic core fusions. The triggering of secular
evolution is found to be linked to the presence of a stellar bulge
so that systems with well-formed stellar bulges do not experience
early gas inflows. Hence, the pace of this morphological loop could 
be regulated by the properties of the galactic central potential wells
which are also affected by the merger history of the objects
(see also Tissera \& Dom\'{\i}nquez-Tenreiro 1998 and Tissera 2000).

We found that the simulated  mass bulge-to-disc
 ratios are within observed range.
It is also noted that during the ODP larger changes are observed in 
the $B/D$ ratios of those objects that experience gas inflows.
The changes during this period are, however, quite disorder.
It is at the fusion of the baryonic clumps that 
changes in the shape parameters are correlated with changes in the
bulge-to-disc ratio. In our simulations, the actual fusions
are responsible of significantly 
 increasing the mass concentration at the centre.
Hence, we found that the fusion of the baryonic cores could be
the process that determine the observed correlation between
the luminosity $B/D$ ratio and the shape parameter or morphological
type.
However, we found no dependence on the relative masses of the
colliding objects.

 Overall, our results indicate that the morphological properties
of galactic objects are the result of their merger histories
within a hierarchical clustering scenario. Based on the good
agreement found so far with observations we support the hypothesis
of mergers as the main morphological driver along the Hubble
sequence. Consequently, the particular and detailed history of 
substructure aggregation could
be a key point in the determination of the astrophysical properties of
galaxies.

\section*{Acknowledgments}

We thank the anonymous referee for useful comments that helped to
improve this paper.
We are grateful to S. Courteau and his collaborators for
making available their observational data prior to publication.
We acknowledge A. Graham and S. Courteau for useful discussions.
We thank Max-Planck Institute for Astrophysics for their hospitality,
where this paper was completed.
This work was partially supported by the
 Consejo Nacional de Investigaciones Cient\'{\i}ficas y T\'ecnicas,
Agencia de Promoci\'on de Ciencia y Tecnolog\'{\i}a and 
 Fundaci\'on Antorchas.


\begin{thebibliography}{}

\bibitem[Aguerri, Balcells \& Peletier 2001]{A01}
Aguerri J.A.L., Balcells M., Peletier R.F., 2001, A\&A, 367, 428 (A01)

\bibitem[Andredakis \& Sanders 1994]{AS94}
Andredakis Y.C., Sanders R.H., 1994, MNRAS,  267, 283

\bibitem[Andredakis, Peletier \& Balcells 1995]{APB95}
Andredakis Y.C., Peletier R.F., Balcells M., 1995, MNRAS,  275, 874 (APB95)

\bibitem{BH96}
Barnes J.E., Hernquist L., 1996, ApJ,  471, 115

\bibitem{ciotti}
Ciotti L., Bertin G., 1999, A\&A, 353, 447

\bibitem{CdJB96}
Courteau S., de Jong R.S., Broeils A.H., 1996, ApJ,  457, L73 (CdJB96) 

\bibitem{dJ96}
de Jong R.S., 1996, A\&A, 313, 45

\bibitem{dalcanton}
Dalcanton J.J., Spergel D.N., Summers F.J., 1997, ApJ, 480, L91

\bibitem{dtts98}
Dom\'{\i}nguez-Tenreiro R., Tissera P.B., S\'aiz A., 
1998, ApJ,  508, L123 (DT98)

\bibitem[Fall 1983]{fall83}
Fall S.M., 1983,  in Athanssoula E., eds, Proc. IAU Symp. 100,
Internal Kinematics and Dynamics of Galaxies, 
Dordrecht, D. Reidel Publishing Co., p. 391

\bibitem{fallefst}
Fall S.M., Efstathiou G., 1980, MNRAS,  193, 189 (FE80)

\bibitem[Gilmore \& Wyse 1998]{gilmore}
Gilmore G., Wyse R.F.G., 1998, AJ, 116, 748

\bibitem{GdB99}
Graham A.W., de Blok W.J.G., 2001, ApJ,  556, 177 (GdB01)

\bibitem{GP99}
Graham A.W., Prieto M., 1999, ApJ,  524, L23 (GP99)

\bibitem[Kauffmann, Guiderdoni \& White 1994]{kauf94}
Kauffmann G.,  Guiderdoni B,  White S.D.M., 1994, MNRAS, 267, 981

\bibitem[Khosroshahi,  Wadadekar \&  Kembhavi  2000]{khosroshahi}
Khosroshahi H.G., Wadadekar Y., Kembhavi A., 2000,
ApJ, 533, 162 (KWK00)

\bibitem{Mac}
MacArthur L.A., Courteau S., Holtzman J.A., ApJ,  accepted (MCH02)

\bibitem{MH94}
Mihos J.C., Hernquist L., 1994, ApJ,  437, L47

\bibitem{MH96}
Mihos J.C., Hernquist L., 1996, ApJ,  464, 641

\bibitem{mmw98}
Mo H.J., Mao S., White S.D.M., MNRAS, 295, 319

\bibitem[Moriondo, Giovanelli \& Haynes 1999]{MGH99}
Moriondo G., Giovanelli R., Haynes M.P., 1999, A\&A, 346, 415 (MGH99)

\bibitem[Navarro \& White 1994]{nw}
Navarro J.F., White S.D.M., 1994, MNRAS, 267, 401

\bibitem[Peebles 1969]{peebles69}
Peebles P.J.E., 1969, ApJ, 155, 393

\bibitem[Pfenniger \& Norman 1990]{pfen90}
Pfenniger D., Norman C., 1990, ApJ, 363, 391

\bibitem{SDTTC01}
S\'aiz A., Dom\'{\i}nguez-Tenreiro R., Tissera P.B., Courteau S.,
2001, MNRAS,  325, 119 (S01)


\bibitem{Sm94}
Steinmetz M., M\"uller E., 1994, A\&A, 281, L97

\bibitem{SW97}
Steinmetz M., White S.D.M., 1997, MNRAS, 288, 545

\bibitem{tiss2000}
Tissera P.B., 2000, ApJ, 534, 636


\bibitem{TDT98}
Tissera P.B., Dom\'{\i}nguez-Tenreiro R., 1998, MNRAS, 297, 177

\bibitem[Tissera, Lambas \& Abadi 1997]{TLA97}
Tissera P.B., Lambas D.G., Abadi M.G., 1997, MNRAS, 286, 384

\bibitem[Tissera et al. 2002]{T02}
Tissera P.B., Dom\'{\i}nguez-Tenreiro R., Scannapieco C.,
S\'aiz A., 2002, MNRAS,  333, 327 (T02)

\bibitem[Trujillo, Graham \& Caon 2001]{trujillo}
Trujillo I., Graham A.W., Caon N., 2001, MNRAS, 326, 869 (TGC01)

\bibitem[van den Bosch 1999]{vandenbosch99}
van den Bosch F.C, 1999, in Carollo C.M., Ferguson H.C., Wyse R.F.G.,
eds, The formation of galactic bulges, 
 Cambridge University Press, p. 50

\bibitem[Weil]{weil}
Weil M.L., Eke V.R., Efstathiou G., 1998, MNRAS, 300, 773.

\end{thebibliography}
\end{document}